\documentclass[intlimits,twoside,a4paper]{article}

\usepackage[cp1251]{inputenc}
\usepackage[eqsecnum]{cmpj3}

\usepackage{bm}

\articletype{A part of the Special collection to the 100th anniversary of the birth of Ihor Yukhnovskii}

\issue{2025}{28}{1}{11601}
\doinumber{10.5488/CMP.28.11601}

\title[Description processes of the interaction of water and aqueous solutions with fuel-containing materials]%
{Description processes of the interaction of water and aqueous solutions with fuel-containing materials in the New Safe Confinement of the ``Shelter'' object}
\author[M. V. Tokarchuk,
        B. M. Markovych, O. S. Zakharyash, O. L. Ivankiv, S.~M.~Mohnyak]
        {M. V. Tokarchuk\orcid{0000-0002-9205-1790}\refaddr{label1,label2},
        B. M. Markovych\orcid{0000-0002-8813-9108}\refaddr{label2}, 
        O. S. Zakharyash\orcid{0000-0002-3083-9478}\refaddr{label2}, 
        O. L. Ivankiv\refaddr{label1}, 
        S.~M.~Mohnyak\orcid{0009-0008-3246-2070}\refaddr{label2}}

\addresses{
\addr{label1} Institute for Condensed Matter Physics of the National Academy of Sciences of Ukraine, 1 Svientsitskii Str., 79011 Lviv, Ukraine
\addr{label2} Lviv Polytechnic National University, 12 S. Bandera Str., 79013 Lviv, Ukraine
}

\date{Received November 15, 2024, in final form February 07, 2025}

\begin{document}

\maketitle

\begin{abstract}
 The main mechanisms and conditions of interaction of lava-like fuel-containing materials (LFCM) with the atmosphere,
 water and aqueous solutions are presented.
 The mechanisms of destruction of the LFCM surface,
 including ion-exchange processes, hydrolysis, dissolution, oxidation, etc., were analyzed. Inhomogeneous diffusion coefficients for UO$_{2}^{2+}$, Cs$^{+}$ ions at the interface ``aqueous solution of radioactive elements -- LFCM'' were calculated.
 The interdiffusion processes and the rate of penetration into the porous medium of the reaction front of the internal hydrolysis of the silicon-oxygen network during interaction with an aqueous solution were analyzed, where 
 the reaction and adsorption processes are hidden in the rate of internal hydrolysis.

\end{abstract}

 \section{Introduction}

 The consequences of the nuclear accidents at the Chornobyl NPP and Fukushima and  their impact on the environment require constant monitoring and research~\cite{Onishi2007,Steinhauser2014800,Yao2014145,Onishi2014372,Konoplev2016568,Taniguchi201912339,Panasiuk201974,Konoplev2020,Fukumoto2020,Kovalenko202347}. Studies \cite{Steinhauser2014800,Konoplev2016568,Konoplev2020,Fukumoto2020} have shown that in all respects the consequences of the accident at the Chernobyl nuclear power plant clearly exceed the consequences of the accident at Fukushima.
 Therefore,
 the issue of the interaction of lava-like fuel-containing materials (LFCM) with the atmosphere,
 water and aqueous solutions is one of the key issues in solving a complex of problems related to both the analysis of the current state of nuclear fuel in various forms in the ``Shelter'' object (SO),
 and in November 2016 in the New Safe Confinement (``Arch'') --- the ``Shelter'' object (NSC-SO),
 as well as with the receipt of long-term predictive estimates for the future.

 It is known that the main sources of water and other liquid components in NSC-SO are the condensation of moisture in the interior of the SO,
 as well as man-made solutions used for dust suppression and control of neutron processes.
 It is important to note that until the introduction of NSC ``Arch'',
 one of the main sources of water in the SO was atmospheric precipitation (rain, snow),
 since the SO was not a hermetic facility.
 Due attention is paid to the problems of water and aqueous solutions in interaction with LFCM in SO both in theoretical and experimental studies,
 starting from the 1990-ies until today~\cite{Yukhnovskii199420,Yukhnovskii199540,Yukhnovskii199619,Klyuchnikov199622,Yukhnovskii1997169,Yukhnovskii1997153,
 Yukhnovskii199763,Borovoy1998104,Bogatov199928,Yukhnovskii1999351,Yukhnovskii1999224,Krinitsyn200021,Yukhnovskii2000361,Shcherbin2000281,
 Yukhnovsky200212,Odintsov2005467,Odintsov202195,Tokarchuk2006319,Odintsov2007467,Yukhnovsky2007112,Chernobyl2007,Glushak2008100112,Panasiuk2010128,
 Panasiuk201974,Kovalenko202347}.
 Water flows are the main means of transporting radionuclides both within the NSC-SO and outside it~\cite{Bogatov199928,Yukhnovskii2000361,Shcherbin2000281,Odintsov2005467,Tokarchuk2006319,Odintsov2007467,Panasiuk2010128,Panasiuk201974,Kovalenko202347},
 as well as an important factor of possible nuclear danger for certain concentrations of LFCM~\cite{Lagunenko201551,Vysotsky201876,Krasnov201922,Vysotskyi202049}.
 In~\cite{Panasiuk201974},
 research was conducted that showed the presence of high concentrations $^{90}$Sr and other radionuclides inside and near the NSC-SO complex,
 which constitutes an environmental hazard associated with the spread of radioactive contamination through groundwater into the environment and particularly  the Pripyat River,
 which drains to the Dnipro basin and is the source of drinking water for the population of Ukraine.

 After the construction of a new safe confinement in 2016 and operation since 2018, there was an urgent need to reassess the state of nuclear and radiation safety of fissile materials~\cite{Vysotsky201876,Krasnov201922,Vysotskyi202049},
 because under the influence of external and internal factors,
 processes are taking place that can increase the intensity of destruction of LFCM and the formation of highly active dust, and 
 lead to the drying up of radioactive water accumulations,
 the exposure of bottom sediments (room 012/7)~\cite{Odintsov202158} and the appearance of radioactive aerosols~\cite{Lagunenko202079,Lagunenko202199,Khan202335},
 as well as changes in the subcritical level (due to a decrease in water concentration) in potentially nuclear-hazardous accumulated LFCM,
 in particular,
 in room 305/2~\cite{Lagunenko201551,Vysotsky201876,Krasnov201922,Vysotskyi202049}. 
 In addition,
 it was experimentally established by the X-ray diffraction method~\cite{Gabielkov201944} that brown,
 black,
 and polychrome LFCM ceramics underwent a partial crystallization with the formation of crystalline phases:
 silicon oxides,
 silicates,
 aluminates,
 ferrites,
 uranates,
 etc.
 The processes of oxidation of uranium oxide UO$_{2}$ in the inclusions,
 which lead to an increase in their volume and to the formation of cracks in the LFCM glass matrix,
 have been confirmed.

 The interaction of LFCM with liquid solutions can initiate a number of physical and chemical processes.
 It is known that the monitoring of the SO atmosphere indicates the presence of nuclear dust of micron and submicron sizes~\cite{Klyuchnikov199622,Borovoy1998104,Baryakhtar199720,Baryakhtar2002449}.
 In rooms with high humidity,
 in which there are accumulations of LFCM,
 the growth of uranium minerals was observed on their surface~\cite{Anderson199344,Kiselev199459,Baryakhtar199818}:
 UO$_{3}\cdot$2H$_{2}$O,
 UO$_{4}\cdot$4H$_{2}$O,
 UO$_{2}$CO$_{3}$,
 UO$_{3}\cdot$1.6O$_{3}\cdot$1.9H$_{2}$O,
 Na$_{4}$UO$_{2}$(CO$_{3}$)$_{3}$,
 in which an important component are crystallization water molecules.
 Experimental studies~\cite{Bogatov199928,Krinitsyn200021,Yukhnovskii2000361} showed that radionuclides are washed out of LFCMs that interact directly with water flows and are transported by water flows to the lower rooms of block B,
 to room 001/3 and through the partition wall to the lower rooms of 3th block. In particular,
 isotopes of uranium $^{235}$U,
 $^{238}$U,
 plutonium $^{238}$Pu,
 $^{239,240}$Pu,
 americium $^{241}$Am,
 cesium $^{134,137}$Cs,
 strontium $^{90}$Sr,
 europium $^{154}$Eu were recorded.
 Moreover,
 the ratios of plutonium/uranium,
 $^{241}$Am/uranium,
 $^{154}$Eu/uranium in the bottom sediments of room 001/3 were close to those known for LFCM,
 and the content of isotopes of cesium $^{134}$Cs,
 $^{137}$Cs and strontium $^{90}$Sr was $4\div7$ times higher than the corresponding values in ``averaged'' fuel.
 Studies of the dynamics of the average concentrations of radionuclides in ``block'' waters allowed us to estimate that in a year in the SO,
 water flows can dissolve and transport several tens of kilograms of uranium and about $10^{14}$ Bq of other radionuclides~\cite{Krinitsyn200021,Shcherbin2000281}.
 In works~\cite{Lagunenko202079,Kalynovskyi202178},
 radioactive aerosols were monitored in the surface layer of the air near the SO in the period 1998--2019.

 The composition of long-lived nuclides in the air of the local area of the SO included nuclides $^{238}$Pu,
 $^{239,240}$Pu,
 $^{241}$Am,
 $^{137}$Cs,
 $^{90}$Sr+$^{90}$Y. These were monitored to assess the radiation safety of personnel.
 In~\cite{Khan202335},
 the radionuclide composition,
 volumetric activities,
 density of radionuclide sediments,
 and dispersion of aerosols collected in 2019--2022 in rooms 012/7, 012/15, 210/7,
 as well as in 2017--2022 in premises 304/3 of the SO,
 where LFCM leaked after the accident.
 The conclusions of this work should be emphasized.
 In particular,
 it established that the median aerodynamic diameter of the carriers of radioactive products of the accident in rooms 012/7, 012/15,
 210/7 has increased significantly over the past 8--10 years and reaches 10~nm.
 This indicates their dispersive origin.
 The presence of transuranic elements in the aerosols in all rooms in ratios that are close to  ratios of bulk LFCM shows that aerosols continue to arise as a result of lava destruction.  It is important to note that in the work \cite{Patsahan20143303}, the destruction processes occurring in the near-surface region of the LPVM were investigated by stochastic computer modelling. The obtained results are qualitatively consistent with the histograms of the sizes of the sprayed particles observed in the lava-like fuel-containing masses that formed during the Chernobyl catastrophe.

 Thus,
 the study of atmospheric and water flows that interact with LFCM,
 as well as aqueous solutions of radionuclides in SO require constant attention,
 since these systems are real sources of uncontrolled migration of radionuclides~\cite{Panasiuk201974}.
 It is also important to note,
 as observations of LFCM~\cite{Klyuchnikov199622,Bogatov199928,Baryakhtar2002449,Anderson199344,Kiselev199459} show, that 
 the surface of LFCM can undergo significant structural changes in the places of interaction with water flows.

 From the point of view of long-term predictive assessments for the future and the strategy of extracting LFCM from SO and safe disposal,
 in our opinion it is still important to rethink the physicochemical processes of LFCM formation.
 Special attention should be paid to the series of works by O.~V.~Zhidkov~\cite{Gonchar200225,Zhidkov200123,Zydkov201186,Zydkov2007443} and Krasnorutsky et al.~\cite{Krasnorutsky201060},
 in which the physico-chemical processes of LFCM formation by melting the uranium fuel  with aggressive silicate melts
 (filling materials of an emergency reactor: serpentinite, sand, concrete, etc.)
 are substantiated. Particularly, the reactions at temperatures of $1050\div1200^\circ$C are relevent to the current degradation processes.
  These differ from the scenarios of LFCM formation presented in works~\cite{Pazukhin199497,Bogatov2008565,Pazukhin200493,Harutyunyan2010240}.
 Another global issue regarding the amount of nuclear fuel in the SO remains open and can be determined definitively after its removal from the SO for safe disposal.
 Today,
 there are many different estimates~\cite{Novoselskij1992129,Kiselev1995252,Kiselev1995306,Checherov2000597},
 which differ significantly from the official data recorded in the IAEA~\cite{GKAE198692}.

 \section{The main mechanisms and conditions of interaction of LFCM with the atmosphere, water and aqueous solutions}

 To understand the processes of interaction of LFCM with the environment,
 it is important to determine the main mechanisms and conditions of interaction of LFCM with the atmosphere,
 water and aqueous solutions.
 Obviously,
 part of the radioactive elements enters the  water streams by leaching from LFCM during interaction with water.
 There is also another way of their being transferred into local waters:
 it is known,
 in particular,
 that the surface of LFCM is constantly destroyed with the formation of nuclear dust containing radionuclides~\cite{Baryakhtar199720,Baryakhtar2002449}.
 In order to suppress dust,
 the internal premises of the SO are periodically sprayed with special solutions.
 Thus,
 dust particles fall into places where water accumulates and
 where the mechanisms of leaching of radionuclides  into aqueous solutions are activated.
 The intensity of interaction of LFCM with the atmosphere,
 water and aqueous solutions is determined both by the physical and chemical parameters of each of the subsystems and by external conditions.
 The main external conditions of such interaction are:
 \begin{itemize}
  \item time of interaction (in SO, the duration extends for decades);
  \item temperature (in the post-accident period, as well as seasonal fluctuations);
  \item humidity (the relative humidity before the installation of NSC ``Arch'' in the premises of the object was due to a durable leaking of  rain~\cite{Bogatov199928}).
        Naturaly, the rain leaking in the SO leads to an increased humidity.
        In addition, an important issue of the current state of the LFCM after the construction of the ``Arch'' and the military actions of the Russian army there are further matters of the state of the ``Arch'' and preservation of its tightness. Ruptured   hermeticity will again lead to the impact of atmospheric precipitation,
        as well as to the possible movement of radioactive dust,
         beyond the borders of NSC ``Arch''.
        The migration of radioactive dust proved to be a significant problem~\cite{Mryglod2005A52,Duviryak2007139,Pavlovskyi202233,Krukovskii202242} in the past;
  \item permanent radiation fields.
 \end{itemize}

 A number of factors influence the course of leaching of radionuclides from LFCM when interacting with aqueous solutions.
 We note among them the following:
 \begin{itemize}
  \item radiation defect formation,
        which as a result of self-irradiation of LFCM leads to the appearance and growth of pores, as well as to structural changes (including the appearance of radiation-disordered regions with characteristic sizes of the order of $20\div$30\,nm~\cite{Baryakhtar199720,Baryakhtar2002449,Baryakhtar199818,Yukhnovsky200212,Patsagan200792,Patsahan20143303,Verkholyak200511,Ignatyuk200964}),
        which increases the effective area of the interaction surface.
        Partial crystallization of brown,
        black and polychrome LFCM~\cite{Gabielkov201944} with the formation of uranium oxides UO$_{2.234}$,
        UO$_{2.34}$,
        U$_{3}$O$_{8}$,
        UO$_{3}$,
        zirconium oxide ZrO$_{2}$ and silicon oxides, silicates, aluminates, ferrites, uranates, often in hydrated form.
        This leads to structural changes of LFCM with the appearance of cracks,
        loss of strength and local destruction.
        At the same time, 
        the effective area of interaction of LFCM with atmospheric flows increases
        (oxidation processes, influence of humidity).
        Predictions of LFCM crystallization were made on the basis of the model studies in work~\cite{Levitskyi2008110};
  \item radiation-induced changes in transfer coefficients,
        in particular diffusion,
        which is important when considering the interphase effects~\cite{Yukhnovsky200212};
  \item oxidation processes of the LFCM surface.
        At the same time,
        reactions play a special role UO$_{2}\rightarrow \mathrm{U}_{3}\mathrm{O}_{7}$/U$_{4}$O$_{9}\rightarrow\mathrm{U}_{3}$O$_{8}$;
  \item  the porous properties of LFCM, in particular, the molecular sieve effect,
        in which up to $10^{21}$ water molecules can be accommodated in 1\,cm$^{3}$ of LFCM.
        That is,
        approximately every tenth silicate ``ring'' or ``silica bridge (O-O) bond'' will be broken~\cite{ArchNo37231998,Zydkov2007443}.
        It is important to note that the porous space of LFCM is a system of gas macro- and micropores connected between themselves and the external environment by microcracks and porous nanochannels ($40\div60$\,nm).
        Gas pores determine the moisture content and heterogeneity of the  LFCM free volume clusters, while microcracks and nanochannels are the mechanisms of volumetric sorption and the speed of water migration in the free volume.
        Water migration processes (retarder, neutron absorber) in the porous space of LFCM clusters are a significant problem,
        because they are one of the main factors in the dynamics of neutron activity of nuclear hazardous LFCM clusters. It is important to note the results of the work~\cite{ Patsahan2007143}, in which the structural and dynamic properties of aqueous uranyl solutions in a silica pore are studied using molecular dynamics methods;
  \item  the physical and chemical properties of aqueous solutions accumulated on the base of the SO
        (``block'' waters) are highly active alkaline-carbonate colloidal solutions with a content CO$_{3}^{2-}$,
        HCO$_{3}^{-}$ (CO$_{3}^{2-}$+H$^{+}\rightarrow \mathrm{HCO}_{3}^{-}$) that varies from zone to zone~\cite{Bogatov199928,Krinitsyn200021,Yukhnovskii1999224,Shcherbin2000281,Yukhnovsky2007112};
  \item the effect of ion transfer processes of radionuclides during the interaction of LFCM with aqueous solutions~\cite{Yukhnovskii1999224,Yukhnovsky2007112};
  \item physical and chemical processes of sediment formation~\cite{Odintsov202158}
        (drying of ``block''  waters),
        their structure,
        chemical composition, temperature
        and radiation exposure.
 \end{itemize}

 \section{Mechanisms of destruction of the LFCM surface}

 The radiation action during the interaction of LFCM with water,
 air or water vapor manifests itself in various corrosive radiolytic products,
 including nitric and carboxylic acids,
 hydrogen peroxide,
 molecular hydrogen and oxygen,
 as well as radicals,
 such as $\mathrm{HO}_{2}^{-}$, $\mathrm{O}_{2}^{-}$.
 These radiolytic products can affect the stability of LFCM due to the changes in the pH and Eh of the solution,
 as well as due to the effects of complex formation involving radionuclides and other components of LFCM~\cite{Davidov1978,Yukhnovskii199763,Tsushima200021,Stasyuk2000743,Tsushima2001365,Mysakovych2002111,Mysakovych2007121,Druchok200511,Druchok2007125}.
 As already noted,
 ``block'' waters are alkaline-carbonate solutions with a relatively low oxidation-reduction potential
 (Eh from minus 100 to plus 100~mV),
 with a variable content of $\mathrm{CO}_{3}^{2-}$ carbonate ions and bicarbonate ions $\mathrm{HCO}_{3}^{-}$,
 $\text{pH}=9\div 12$~\cite{Bogatov199928,Krinitsyn200021}.
 In such aqueous solutions,
 uranium is in the form of $\mathrm{U}^{6+}$ and $\mathrm{U}^{4+}$
 (pair potential $\mathrm{U}^{4+}/\mathrm{UO}_{2}^{2+}=0.334$\,V).
 The uranyl ion $\mathrm{UO}_{2}^{2+}$ at $\mathrm{pH}=8\div10$ and an excess of carbonate ions $\mathrm{CO}_{3}^{2-}$ forms the complex $[\mathrm{UO}_{2} (\mathrm{CO}_{3})_{3}]^{4-}$,
 and at $\mathrm{pH}=6\div8$ the complex $[\mathrm{UO}_{2} (\mathrm{CO}_{3})_{2}]^{2-}$ dominates.
 Uranium $\mathrm{U}^{4+}$ probably forms the complex $[\mathrm{U}(\mathrm{CO}_{3})_{3}]^{2-}$,
 although 
 the formation of mixed uranium hydroxocarbonate $[\mathrm{U}(\mathrm{OH})_{2} (\mathrm{CO}_{3})_{2}]^{2-}$ is possible.
 Carbonates and hydrocarbonate complexes containing $\mathrm{U}^{4+}$, $\mathrm{U}^{6+}$ are transported by water flows to the lower levels of the block,
 where under certain conditions,
 in particular,
 when water evaporates,
 their crystallization may occur with the formation of solid sediments~\cite{Odintsov202158}.
 Plutonium ions $\mathrm{Pu}^{4+}$ in alkaline-carbonate solutions at $\mathrm{pH}=7\div9$ form a neutral complex of plutonium dicarbonate $\mathrm{Pu}(\mathrm{CO}_{3})_{2}$,
 and in the region up to $\mathrm{pH}=10$ --- a complex of plutonium tricarbonate $\mathrm{Pu}(\mathrm{CO}_{3})_{3}^{2-}$.
 Most of the $\mathrm{PuO}_{2}^{2+}$ ions,
 as a result of hydrolysis,
 the effects of complex formations and polymerization,
 take part in the formation of large colloidal aggregates~\cite{Yukhnovskii199763,Davidov1978,Tsushima200021,Stasyuk2000743}.

 Given the glassy composite porous structure of LFCM,
 the following corrosion mechanisms can be identified when interacting with water and aqueous solutions:

 1. Ion exchange and diffusion.

 When porous LFCM  interacts with aqueous solutions,
 water molecules,
 as well as $\mathrm{H}^{+}$,
 $\mathrm{OH}^{-}$,
 $\mathrm{H}_{3}\mathrm{O}^{+}$ ions diffuse into the LFCM matrix,
 which is accompanied by ion exchange processes:
 $\mathrm{H}^{+}$,
 $\mathrm{H}_{3}\mathrm{O}^{+}$ ions replace $\mathrm{Na}^{+}$,
 $\mathrm{Cs}^{+}$ ions in the silicon-oxygen structure etc.,
 which pass into the solution
 \begin{equation}  \label{eq.01}
  [n(\mathrm{Si}-\mathrm{O}^{-})-\mathrm{X}^{n}]+[[n \mathrm{H}_{3}\mathrm{O}^{+}]]
  \rightarrow
  [n(\mathrm{Si} - \mathrm{OH})] +  [[\mathrm{X}^{n}]] + [[n\mathrm{H}_{2}\mathrm{O}]],
 \end{equation}
 where $[\ldots]$ and $[[\ldots]]$ indicate the presence of the corresponding chemical compound in the solid phase
 (on the surface) or in solution;
 soluble elements $\mathrm{X}=[[\mathrm{K}^{+},\mathrm{Na}^{+},\mathrm{Cs}^{+}, \mathrm{etc}.]]$ diffusing from the vitreous surface of LFCM are replaced by hydrogen.
 At the same time,
 the pH of the solution increases,
 that is,
 the concentration of hydroxides increases,
 which,
 in turn,
 interact with silicon-oxygen bonds $\equiv \mathrm{Si}-\mathrm{O}-\mathrm{Si}\equiv$ according to the hydrolysis scheme.
 At the same time,
 a near-surface diffusion layer with an increased reaction activity is formed.

 2. Hydrolysis of silicon-oxygen bonds leads to processes of depolymerization of the LFCM surface:
 \begin{equation} \label{eq.02}
  [\equiv \mathrm{Si} - \mathrm{O} - \mathrm{Si}\equiv]+ [[\mathrm{OH}^{-}]]
  \leftrightarrow
  [\mathrm{Si} - \mathrm{O}^{-} - \mathrm{Si} - \mathrm{OH}],
  [\equiv \mathrm{Si} - \mathrm{O} - \mathrm{Si}\equiv]+ [[\mathrm{H}_{2}\mathrm{O}]]\leftrightarrow 2[\equiv \mathrm{Si} - \mathrm{O} - \mathrm{H}].
 \end{equation}
 As can be seen,
 the hydrolysis reaction can occur without releasing silicon into the solution.
 At the same time,
 the silicon-oxygen bonds are broken,
 the $\mathrm{OH}^{-}$ ion joins them,
 and the hydrogen ion quickly migrates along the structure and,
 capturing an electron,
 turns into atomic hydrogen.

 3. Dissolution of the vitreous matrix  LFCM.

 There is also another version of the reaction,
 when the hydrolysis of silicon-oxygen bonds occurs with the release of silicon into the solution:
 \begin{align} \label{eq.03}
  [\equiv \mathrm{Si} - \mathrm{O} - \mathrm{Si}\equiv] &+ [[\mathrm{OH}^{-}]]
  \leftrightarrow
  [[\mathrm{Si} - \mathrm{O}^{-} - \mathrm{Si} - \mathrm{OH}]], \nonumber\\
  [\equiv \mathrm{Si} - \mathrm{O} - \mathrm{Si}\equiv] &+ [[\mathrm{H}_{2}\mathrm{O}]]
  \leftrightarrow 2[[\equiv \mathrm{Si} - \mathrm{O} - \mathrm{H}]], \nonumber\\
  [\equiv \mathrm{Si} - \mathrm{O} - \mathrm{Si}\equiv] &+ 4[[\mathrm{OH}^{-}]]
    \leftrightarrow
    [\mathrm{Si} - \mathrm{O}^{-}] + [\mathrm{Si} - \mathrm{OH}_{4}].
 \end{align}
 At the same time,
 dissolution of the matrix material occurs.
 It is this process,
 as shown by experimental studies for glasses~\cite{Sobolev1999240,Belustin198395},
 that should be dominant at high pH values.
 It is important to note that the interdependent processes of ion exchange and matrix dissolution lead to the transformation of part of the diffusion layer into  transition layer.
 When the $\equiv \mathrm{Si} - \mathrm{O} - \mathrm{Si}\equiv$ silicate matrix is dissolved,
 chemical elements,
 as well as entire complexes,
 pass into the diffusion layer,
 which leads to the formation of a certain amount of $\mathrm{SiO}_{2}$ at the ``transition layer -- diffusion layer'' interface,
 which can enter the solution.
 Due to the surface destruction of silicon-oxygen bonds,
 in addition to $\mathrm{Si}-\mathrm{O}$,
 $\mathrm{Si}-\mathrm{OH}$,
 $\mathrm{Si}(\mathrm{OH})_{4}$,
 $\mathrm{Si}(\mathrm{OH})_{6}^{2-}$ (hydrolysis products),
 $\mathrm{Cs}^{+}$ ions also gradually pass into the diffusion and transition  layers and further into the solution,
 $\mathrm{Sr}^{2+}$,
 $\mathrm{UO}_{2}^{2+}$,
 $\mathrm{PuO}_{2}^{2+}$,
 which are hydrolyzed in the solution with the subsequent formation of complexes.
 The dissolution of the silicate matrix occurs selectively,
 since in the process of dissolution in the layer individual components have different activity.

 In addition to the above-discussed mechanisms of dissolution of the silicon-oxygen network of LFCM,
 which are associated with $\mathrm{H}^{+}$,
 $\mathrm{OH}^{-}$,
 $\mathrm{H}_{3}\mathrm{O}^{+}$ ions,
 it is also necessary to take into account the dissolution mechanisms associated with the existence in LFCM,
 on the one hand,
 of higher oxide forms of uranium $\mathrm{UO}_{3}$,
 $\mathrm{U}_{3}\mathrm{O}_{8}$, etc. On the other hand,
 the presence of bicarbonate ions $\mathrm{HCO}_{3}^{-}$,
 as well as molecular oxygen $\mathrm{O}_{2}$ in highly active ``block'' aqueous solutions.
 At the same time,
 a reaction may occur on the surface of LFCM
 \begin{equation} \label{eq.04}
  2[\mathrm{U}_{3}\mathrm{O}_{8}]+3x[[\mathrm{H}_{2}\mathrm{O}]]
  \longrightarrow
  [\mathrm{U}_{3}\mathrm{O}_{7}]+3[\mathrm{UO}_{3}\cdot x\mathrm{H}_{2}\mathrm{O}],
 \end{equation}
 and the oxidation processes of $\mathrm{UO}_{2}$ on the surface of LFCM lead to the formation of $\mathrm{UO}_{3}$:
 \begin{equation} \label{eq.05}
  [\mathrm{UO}_{2}]+ \frac{1}{2}[[\mathrm{O}_{2}]]
  \rightleftarrows^{K_{1}}_{K_{- 1}} [\mathrm{UO}_{3}].
 \end{equation}
 Interaction on the surface of $\mathrm{UO}_{3}$ with bicarbonate ions $\mathrm{HCO}_{3}^{-}$ according to the scheme~\cite{Pablo19993097}
 \begin{equation} \label{eq.06}
  [\mathrm{UO}_{3}]+ [[\mathrm{HCO}_{3}^{-}]] \rightarrow^{K_{2}}[\mathrm{UO}_{3}-\mathrm{HCO}_{3}^{-}]
 \end{equation}
 stimulates the $\mathrm{UO}_{3}-\mathrm{HCO}_{3}^{-}$ complex to separate into the solution
 \begin{equation} \label{eq.07}
  [\mathrm{UO}_{3}-\mathrm{HCO}_{3}^{-}] \rightarrow^{K_{3}}[[\mathrm{U}(\mathrm{VI})]].
 \end{equation}
 The yield coefficient of the dissolution reaction can be determined from the ratio:
 \begin{equation} \label{eq.08}
  r=\frac{\rd}{\rd t}[[\mathrm{U}(\mathrm{VI})]]=K_{3}[\mathrm{UO}_{3}-\mathrm{HCO}_{3}^{-}],
 \end{equation}
  and, considering all the stages from oxidation to dissolution, taking into account the condition of process stationarity
 $K_{3}[\mathrm{UO}_{3}-\mathrm{HCO}_{3}^{-}]=K_{2}[\mathrm{UO}_{3}][[\mathrm{HCO}_{3}^{-}]]$, we have

 \begin{equation} \label{eq.09}
  r=\frac{K_{1}K_{2}[\mathrm{UO}_{2}]_{\mathrm{tot}}[[\mathrm{O}_{2}]][[\mathrm{HCO}_{3}^{-}]]}{K_{-1}+K_{2}[[\mathrm{HCO}_{3}^{-}]]+K_{1}[[\mathrm{O}_{2}]]}.
 \end{equation}
 Reaction constants (\ref{eq.05})--(\ref{eq.07}) can be determined experimentally for ``block'' aqueous solutions,
 which is essential for elucidating the contribution of such a mechanism to the leaching of uranium isotopes from the surface of LFCM.
 The exit of the complex into the solution together with the processes of hydration~(\ref{eq.04}) and oxidation~(\ref{eq.05}) can be considered as the process of dissolution of LFCM.

 4. Repolymerization.
 In the transition layer,
 the process of repolymerization can occur,
 which is associated with the rearrangement of the hydrolyzed structures of $\mathrm{Si}-\mathrm{OH}$, $\mathrm{Si}(\mathrm{OH})_{4}$,
 $\mathrm{Si}(\mathrm{OH})_{6}^{2-}$ and the formation of $\mathrm{Si}-\mathrm{O}-\mathrm{Si}$ nanocrystals with the release of $\mathrm{H}_{2}\mathrm{O}$  water molecules~\cite{Sobolev1999240}.

 5. Sorption and formation of colloids.
 Sedimentation.
 In the process of dissolution of the silicon-oxygen structure of LFCM and the release of individual chemical elements or the whole complexes into the transition layer,
 and then into the solution,
 the chemical composition of the solution itself changes,
 in which,
 in turn,
 the processes of sorption and formation of colloids can occur.
 In particular,
 sorption of uranium can occur according to the scheme
 \begin{equation} \label{eq.010}
  2(\mathrm{HO}-\mathrm{Si}\equiv)+[\mathrm{UO}_{2}(\mathrm{CO}_{3})_{3}]^{4-}]
  \leftrightarrow
  \mathrm{UO}_{2}(\mathrm{O}-\mathrm{Si}\equiv)_{2}+3\mathrm{CO}_{3}^{2-}+2\mathrm{H}^{+},
 \end{equation}
 \begin{equation} \label{eq.011}
  \mathrm{UO}_{2}^{2+}+2(\mathrm{HO}-\mathrm{Si}\equiv)
  \leftrightarrow
  \mathrm{UO}_{2}(\mathrm{O}-\mathrm{Si}\equiv)_{2}+2\mathrm{H}^{+},
 \end{equation}
 with the release of hydrogen ions,
 which in the solution are hydrolyzed to $\mathrm{H}_{3}\mathrm{O}^{+}$ and can further participate in ion-exchange diffusion processes on the surface of LFCM.
 Formation of colloids,
 in particular $\mathrm{Pu}(\mathrm{CO}_{3})_{2}$,
 $[\mathrm{Pu}(\mathrm{CO}_{3})_{3}]^{2-}$,
 $[\mathrm{UO}_{2}(\mathrm{CO}_{3})_{3}]^{4-}$,
 $[(\mathrm{UO}_{2})_{2}\mathrm{CO}_{3}]^{2+}$,
 $[\mathrm{U}(\mathrm{CO}_{3})_{3}]^{2-}$,
 $[\mathrm{U}(\mathrm{OH})_{2}(\mathrm{CO}_{3})_{2}]^{2-}$,
that are characteristic of ``block'' waters~\cite{Bogatov199928},
 significantly affects the pH of the solution,
 the content of which changes up to saturation with dissolved components of LFCM,
 as a result of which they precipitate.
 The deposited layer can consist of both amorphous and crystalline phases containing $\mathrm{Pu}(\mathrm{OH})_{4}$,
 $\mathrm{UO}_{2}\mathrm{CO}_{3}$,
 $\mathrm{Na}_{4}\mathrm{UO}_{2}(\mathrm{CO}_{3})_{3}$,
 which contain elements from LFCM and aqueous solutions.
 Obviously,
 such transformations lead to the formation of uranium minerals UO$_3\cdot$2H$_2$O, UO$_4\cdot$4H$_2$O, 
 $\mathrm{UO}_{2}\mathrm{CO}_{3}$,
 $\mathrm{Na}_{4}\mathrm{UO}_{2}(\mathrm{CO}_{3})_{3}$,
 which were formed on the surface of LFCM.

 6. Weathering.
 When vitreous LFCMs are exposed to atmospheric flows containing water vapor and other reactive gases
 (for example, $\mathrm{CO}_{2}$),
 as well as during short-term contact with aqueous solutions,
 in particular during dust suppression,
 weathering processes occur,
 which are accompanied by reactions~(\ref{eq.01})--(\ref{eq.011}) with subsequent evaporation of moisture.

 Thus,
 under the influence of aqueous alkaline-carbonate solutions with $\mathrm{pH}=9\div11$ and under the influence of constant radiation in LFCM as a result of complex physicochemical processes
 (hydrolysis, dissolution of silicon-oxygen bonds of the surface of LFCM, cesium is leached from their ultramicropores and micropores, strontium, uranium, plutonium, americium, etc.)
 silicon-oxygen bonds can be destroyed,
 that is,
 the corresponding silicate bonds dissolve.
 These processes can also affect the mechanical,
 dielectric,
 and magnetic properties of LFCM~\cite{ArchNo37241998,Gonchar1996173,Zhidkov20006,Moina200115}.
 Depletion of the silicon-oxygen matrix by its heavy components is confirmed by experimental studies~\cite{Bogatov199928,Krinitsyn200021,Shcherbin2000281}.
 To predict such complex and interdependent phenomena,
 it is necessary to conduct a complex of experimental and theoretical studies.
 Some computational and theoretical evaluations of specific processes will be given in the following sections.

 \section{Calculation and theoretical assessment of the main parameters of interaction of LFCM with the environment}

 It is known that LFCMs have a porosity of the order of $3\div10$ percent with transverse dimensions up to 5\,nm depending on the type of LFCM~\cite{Tokarchuk2006319}.
 Therefore,
 when studying the leaching of radionuclides from LFCM,
 their adsorption on the surface of LFCM~\cite{Stasyuk2007132},
 the effects of adhesion,
 diffusion of radionuclides from the volume of LFCM to the surface,
 it is necessary to take into account the porosity of these materials.
 Note that in a porous medium that is in contact with a liquid solution,
 its properties may change significantly when the solution  and any colloids penetrate the pores.
 At the same time,
 due to spatial limitations,
 the main thermodynamic and kinetic characteristics of the solution itself also change.
 These circumstances must be taken into account when modelling the behavior of LFCM.

 In the leaching process,
 the water in contact with LFCM is a weakly concentrated solution of the electrolyte,
 which contains cations $\mathrm{PuO}_{2}^{2+}$,
 $\mathrm{UO}_{2}^{2+}$,
 $\mathrm{AmO}_{2}^{2+}$,
 $\mathrm{Cs}^{+}$,
 $\mathrm{Sr}^{2+}$,
 $\mathrm{H}^{+}$,
 $\mathrm{Ca}^{2+}$,
 $\mathrm{Na}^{+}$, etc.,
 anions $\mathrm{OH}^{-}$,
 $\mathrm{CO}_{3}^{2-}$,
 $\mathrm{HCO}_{3}^{-}$, etc.,
 and as well as various hydrate complexes and radiolysis products due to exposure to $\alpha-\beta-\gamma$ irradiation.
 Experimental studies of concentration,
 sorption and other surface properties during the interaction of LFCM with aqueous solutions of radioactive elements in the ``Shelter'' facility is a complex technological task.
 Therefore,
 there is a need to build a physical model close to reality that would describe the surface properties of highly active alkaline-carbonate aqueous solutions in contact with the silicate surface of LFCM.

 In order to study the influence of the aqueous solution on the properties of the porous matrix during the solution sorption,
 as well as changes in the properties of the solution in the pores,
 we considered the model of the electrolyte solution in contact with the porous medium, which is based on works~\cite{Holovko199934,Holovko2000391}.
 The electrolyte is considered as a system of charged particles of various types
 ($\mathrm{UO}_{2}^{2+}$, $\mathrm{AmO}_{2}^{2+}$, $\mathrm{Cs}^{+}$, $\mathrm{Sr}^{2+}$ and $\mathrm{OH}^{-}$, $\mathrm{CO}_{3}^{2-}$, $\mathrm{HCO}_{3}^{-}$)
 in the environment of dipoles simulating the molecules with the dielectric constant of water
 ($\varepsilon_{\mathrm{H}_{2}\mathrm{O}} = 81$).
 The porous matrix is modelled by a set of fixed ions and dipoles, and
 its dielectric constant is selected depending on the type of LFCM within ($\varepsilon_{m} = 5\div15$).
 The interaction between particles has a long-range electrostatic character.
 Particle sizes and their close-range interactions are effectively taken into account.
 On the basis of this model and work~\cite{Yukhnovskii2000361},
 heterogeneous diffusion coefficients for $\mathrm{UO}_{2}^{2+}$,
 $\mathrm{Cs}^{+}$ ions at the interface ``aqueous solution of radioactive elements -- LFCM'' were calculated.
For this purpose, the expression for the inhomogeneous diffusion coefficient of ions was used, which was obtained in detail in the work ~\cite{Yukhnovskii2000361}:
 \[
 D^{\alpha}(\vec{r})=\frac{k_{\rm B}T}{m_{\alpha}}\sqrt{\frac{\piup}{2\bar{\lambda}_{2}^{\alpha\alpha}(\vec{r})}},
 \]
where $k_{\rm B}$ is the Boltzmann constant, $T$ is the equilibrium temperature, $m_{\alpha}$ is the mass of ions of the sort $\alpha$, $\vec{r}$ is the radius vector of the spatial coordinate of the particle, $\bar{\lambda}_{2}^{ \alpha\alpha}(\vec{r})$ is the renormalized second moment through the zeroth moment of the time correlation function of the momentum densities of the corresponding ion of the sorts $\alpha$ and $\gamma$, the expression for which is given in the Appendix.

 \begin{figure}[hbtp]
  \centering
  \includegraphics[width=0.5\textwidth]{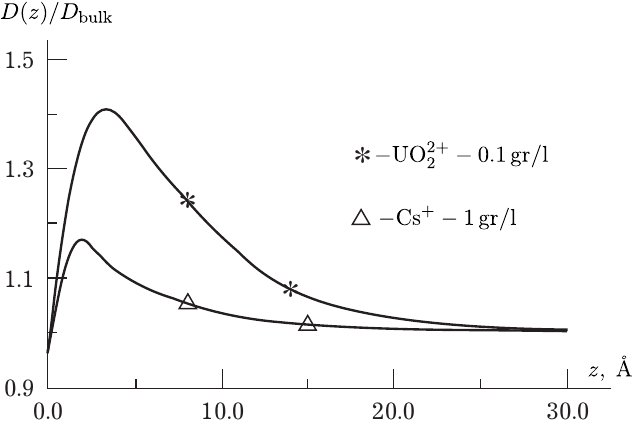}\\
  \caption{Self-diffusion coefficients for UO$_{2}^{2+}$,
           Cs$^{+}$ ions in the near-surface layer of the ``LFCM surface -- aqueous solutions'' system.}\label{Fig1}
 \end{figure}

 \begin{figure}[hbtp]
  \centering
  \includegraphics[width=0.5\textwidth]{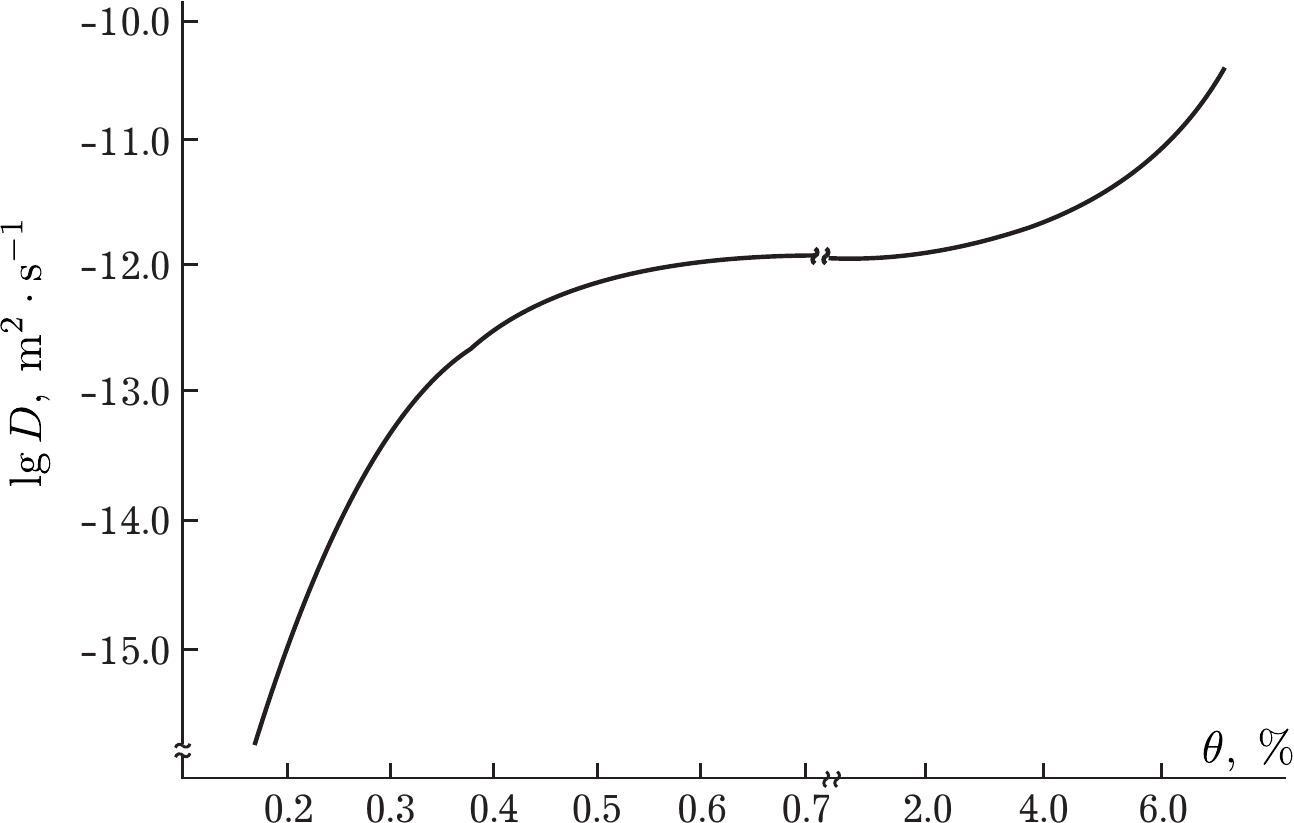}\\
  \caption{Dependence of the uranium diffusion coefficient on the relative water content in a sample of model ceramics at 20$^\circ$C.}\label{Fig2}
 \end{figure}

 As we can see from figure~\ref{Fig1},
 heterogeneous diffusion coefficients,
 normalized to volume values,
 in the near-surface layer of the system ``LFCM surface -- aqueous solutions'' have a region of rapid diffusion of ions in the direction from the surface into the aqueous solution of radioactive elements.

 In order to assess the influence of atmospheric humidity when interacting with LFCM,
 using the phenomenological theory of diffusion in porous media~\cite{Prokhorov198197}, qualitative estimates of the diffusion coefficient of uranium in model porous ceramics with changes in humidity were made.
 The results are presented in figure~\ref{Fig2}.
 We see that at low relative humidity,
 the diffusion coefficient of uranium increases,
 and within the range of $2\div3$ percent it remains almost constant.
 After that,
 it begins to grow gradually,
 which can be explained by the inclusion of other leaching mechanisms.
 The mechanisms of diffusion growth both in the range from $0.2$ to $0.7$ percent,
 and in the range from $2$ percent and further are insufficiently studied.
 At a low relative humidity,
 we assumed that the water in the pores is in a gaseous state and for the calculation we used expressions for the diffusion coefficient for gas in porous media~\cite{Yukhnovskii2000361}.
 In the region of 2 percent and higher values of relative humidity,
 the behavior of the diffusion coefficient under the condition of considering water as a liquid.
 However,
 the microscopic mechanisms of such behavior of radionuclide diffusion coefficients in glassy systems with changes in humidity remain unclear.

 It is also important to note the fact that in studies of vitreous systems (multi-component glasses, ceramics)~\cite{Ivanov1991191,Ivanov1986398,Bellanger199783}  as promising materials for burying radioactive waste,
 it was established that humidity has an extremely strong effect on the diffusion of uranium,
 neptunium,
 americium,
 and plutonium.
 In particular,
 work~\cite{Ivanov1986398} shows that in dry samples at 20$^\circ$C the value of the diffusion coefficient of $^{237}\mathrm{Np}$ is below the sensitivity limit of the used measurement method.
 However,
 in the range of changes in relative humidity from $0.17$ to $0.5$ percent,
 the diffusion of this radionuclide increases from $10^{-16}$ to $10^{-12}$\,m$^{2}\cdot$s$^{-1}$ and with a further increase in humidity (from $0.7$ percent and above) it remains practically constant.

 \section{Computational and theoretical evaluation of the dissolution of the silicon-oxygen network of LFCM as a factor of surface destruction: features of the interaction with alkaline-carbonate solutions of radioactive elements}

 Quantitatively and qualitatively,
 interdiffusion processes in the diffusion layer of reactions~(\ref{eq.01})--(\ref{eq.03}) can be described by diffusion equations~\cite{Belustin198395,Kurylyak200212,Yukhnovsky2007112,Glushak2008100112}:
 \begin{equation} \label{eq.012}
  \frac{\partial }{\partial t}n_{i}
  =
  \frac{\partial }{\partial x}\left(D_{ij}\frac{\partial }{\partial x}n_{j}\right)
  -\frac{\partial }{\partial x}(v_{h}n_{i}),
 \end{equation}
 where $n_{i}$ is the molar fraction of the ion of type $i$ ($i = 1,2$),
 $1$ is the ion from the LFCM,
 and $2$ is the ion from the solution region.
 $D_{12}(x,n_{i})$ is the mutual diffusion coefficient;
 $t$ is time;
 $x$ is coordinate;
 $v_{h}(x, t)$ is the rate of penetration into the porous medium of the front of the internal hydrolysis reaction~(\ref{eq.02}),
 which accompanies the diffusion of water.
 That is,
 the total process in the reaction layer is considered as interdiffusion of ions with internal hydrolysis of the silicon-oxygen network of LFCM.
 In the general case,
 $D_{ij}(x,n_{i})$ depends both on the nature of the interaction of ions and on the structural distribution of ions,
 solution molecules and the structure of the porous medium.
 For the limiting case,
 when $D_{ij}=D$ does not depend on the spatial coordinates $x$ and $v_h$ and is considered as a constant value,
 as well as taking into account the boundary conditions of the first kind for the number of ions of type $1$ leached during time $t$ from $1$\,cm$^{2}$ of the LFCM surface,
 from the equation~(\ref{eq.012}) for the dimensionless quantity $Q_{z,1}= Q_{t,1}/(2c_{10} D/v_{h})$ we obtain~\cite{Belustin198395,Kurylyak200212}
 \begin{equation} \label{eq.013}
  Q_{z,1}=\left(z^{2}+\frac{1}{2}\right)\mathrm{erf}(z)+z^{2}+\piup^{-1/2}\,\re^{-z^{2}},
 \end{equation}
 where $z=v_{h}/[2(D/t)^{1/2}]$ is a dimensionless value,
 and $c_{10}$ is the initial concentration (mol/cm$^{3}$) of ion of type 1 in LFCM.
 For different intervals of the change of parameter $z$,
 equation~(\ref{eq.013}) can be approximated by expressions:
 \begin{eqnarray} \label{eq.014}
  Q_{z,1}=2z/\piup^{-1/2},\quad
  Q_{t,1}=2c_{10} \,\piup^{-1/2}(Dt)^{1/2},  \\
  0<z=v_{h}\big/\left[2(D/t)^{1/2}\right]\leqslant 0.03, \nonumber
 \end{eqnarray}
 \begin{eqnarray} \label{eq.015}
  Q_{z,1}=2z/\piup^{-1/2}(1+z),\quad
  Q_{t,1}=c_{10}\,\piup^{-1/2}\left[2(Dt)^{1/2}+v_{h}t\right] , \\
  0.03\leqslant z=v_{h}\big/\left[2(D/t)^{1/2}\right]< 0.7,  \nonumber
 \end{eqnarray}
 \begin{eqnarray} \label{eq.016}
  Q_{z,1}=2z^{2}+ 1/2,\quad  Q_{t,1}=c_{10}(D/v_{h} + v_{h}t),\\
  0.7\leqslant z=v_{h}\big/\left[2(D/t)^{1/2}\right]. \nonumber
 \end{eqnarray}

 \begin{figure}[hbtp]
  \centering
  \includegraphics[width=0.5\textwidth]{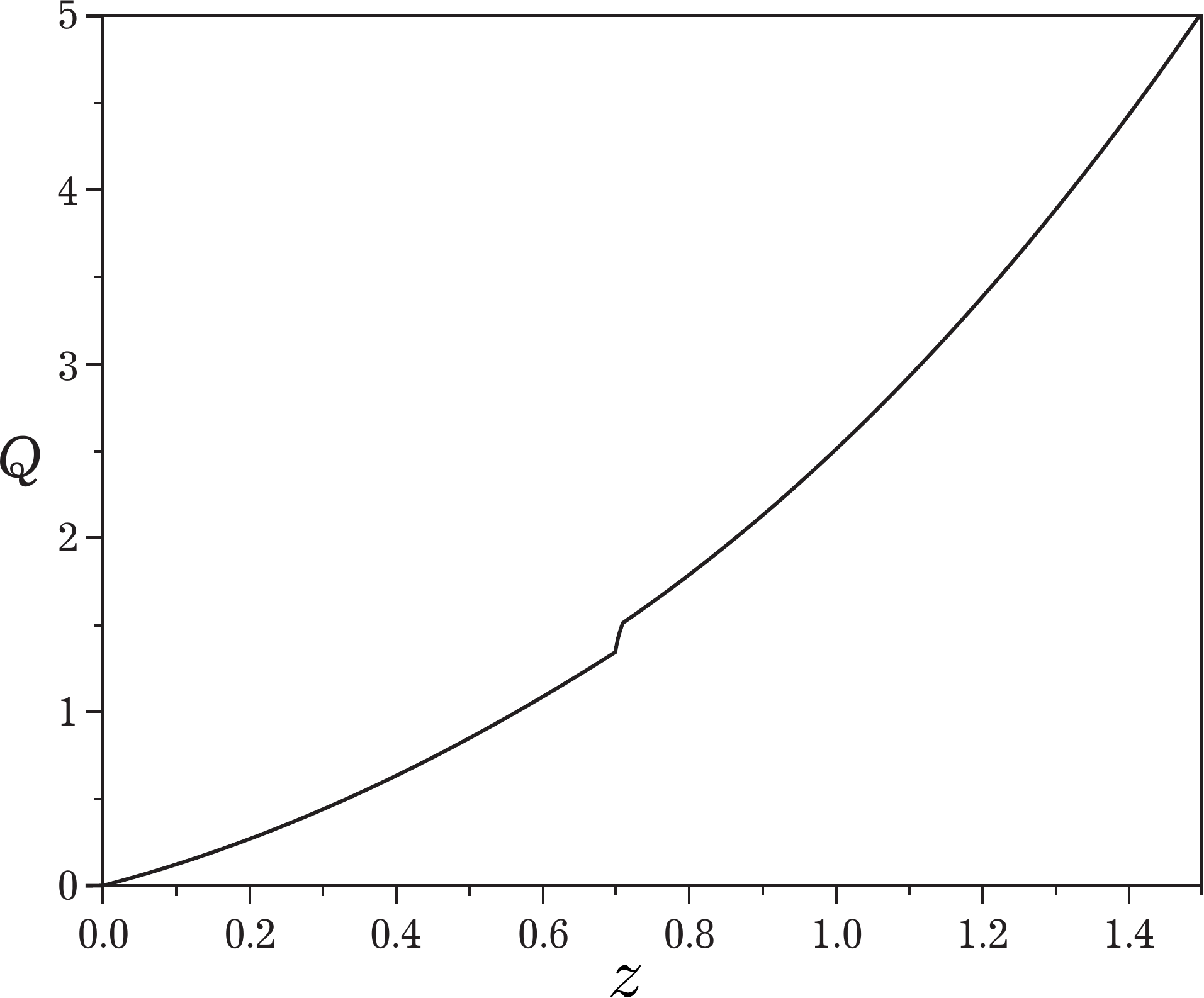}
  \caption{Dependence of the number of type 1 ions leached from the surface of LFCM on the parameter~$z$.}\label{fig3}
 \end{figure}

 In figure~\ref{fig3} shows the dependence of the quantity $Q_{z,1}$ for ions of type 1 on $z$.
 As we can see, a 
 monotonous growth is observed in the range of $z$ changes from $0$ to $1.5$.
 Since $z=v_{h}/[2(D/t)^{1/2}]$,
 the leaching of grade 1 ions from the porous matrix increases under the condition $v_{h} \gg (D/t)^{1/2}$,
 that is,
 when the area of penetration into the porous matrix of the reaction front of internal hydrolysis~(\ref{eq.02}) will increase faster than the average diffusion distance traveled by an ion in a porous medium.
 In the general case,
 the function $v_{h}(x, t)$ is mutually related to $D_{ij}(x, n_{i})$ and the nature of this relationship depends on the nature of the interaction and structural distribution of ions and molecules in both the porous medium and the aqueous solution.
 At the same time,
 the conducted evaluations show that the proposed approach based on model~(\ref{eq.012}) is promising and can be used to obtain quantitative estimates for LFCM.

 In the work~\cite{Report2004},
 experimental studies were carried out in laboratory conditions of the processes of leaching $\beta$ and $\alpha$-active radionuclides from three types of LFCM:
 black,
 polychrome and brown when interacting with water.
 It was shown that the leaching of $\beta$-active elements ($^{90}\mathrm{Sr}$) from LFCM occurs faster than transuranium ones.
 It is important to note that the conditions of the LFCM leaching experiment in laboratory studies are significantly different from real conditions.
 Aqueous solutions that interacted  with LFCM in SO are alkaline ($\mathrm{pH}=9\div11$),
 and double distilled water with $\mathrm{pH}=7$ was used in the experiment.
 Sampling of block water~\cite{Krinitsyn200021,Shcherbin2000281,Tokarchuk2006319} in certain rooms recorded rather high concentration of transuranics,
 not to mention isotopes $\mathrm{Cs}$, $\mathrm{Sr}$ and other radionuclides.
 The diffusion model considered by us qualitatively reproduces the experimental results. For a detailed study of the leaching processes, in particular, ions $\mathrm{Cs}^{+}$, $\mathrm{Sr}^{2+}$, $\mathrm{UO}_{2}^{2+}$ and other nuclides from LFCM when interacting with aqueous solutions, it is necessary to calculate the flow-flow time correlation functions for the components, which are related to the mutual diffusion coefficients $D_{ij}$ by the Kubo ratio.

\section{Result}

 We analyzed the main mechanisms and conditions of interaction of LFCM with the atmosphere, water and aqueous solutions in the conditions of NSC-SO.
 Computational and theoretical evaluations of the influence of the aqueous solution on the properties in the porous matrix during solution sorption showed that:
 \begin{itemize}
   \item in the near-surface region of the volume phase ``solution -- LFCM surface'' heterogeneous diffusion coefficients of $\mathrm{UO}_{2}^{2+}$, $\mathrm{Cs}^{+}$ ions have a region of fast diffusion,
       which is associated with electro-polarization effects;
   \item atmospheric humidity interacting with LFCM can increase the uranium diffusion coefficient by several orders of magnitude with increasing relative humidity;
   \item the diffusion model describing the interdiffusion processes and the rate of penetration into the porous medium of the reaction front of the internal hydrolysis of the silicon-oxygen network during interaction with an aqueous solution is considered. On its basis, estimates of the release of ions from the matrix were obtained depending on the ratio of interdiffusion coefficients, the rate of internal hydrolysis and the interaction time. The diffusion model of leaching must be generalized taking into account the peculiarities of the interaction of LFCM with aqueous solutions.
       In this case, in particular, it is necessary to take into account the fact that due to the negative charge of the LFCM surface, the hydrogen ions formed during radiolysis processes in the near-surface region will quickly be electrostatically attracted to the LFCM surface and thereby activate the ion exchange mechanisms for the release of radionuclides from the LFCM matrix into the solution.
 \end{itemize}

\section{Discussion}

 Calculation-theoretical estimates of the influence of the aqueous solution on the properties in the porous matrix during solution sorption are qualitative,
 as they are based on a simplified ion-dipole model that takes into account the electropolarization nature of the interactions. This model does not take into account essential electronic processes,
 which are necessary for calculation and theoretical estimations of electrical conductivity and thermal conductivity.
 In the proposed diffusion model,
 which describes the interdiffusion processes and the rate of penetration into the porous medium of the reaction front of the internal hydrolysis of the silicon-oxygen mesh when interacting with an aqueous solution,
 the reaction and adsorption processes are hidden in the rate of internal hydrolysis.
 These terms are known theoretically,
 but reaction constants,
 in particular~(\ref{eq.01})--(\ref{eq.011}),
 are necessary for calculation,
 and considerable difficulties arise in their theoretical calculation,
 and therefore they are usually determined experimentally.

 \begin{figure}[hbt]
  \centering
  \includegraphics[clip=true,width=0.75\textwidth]{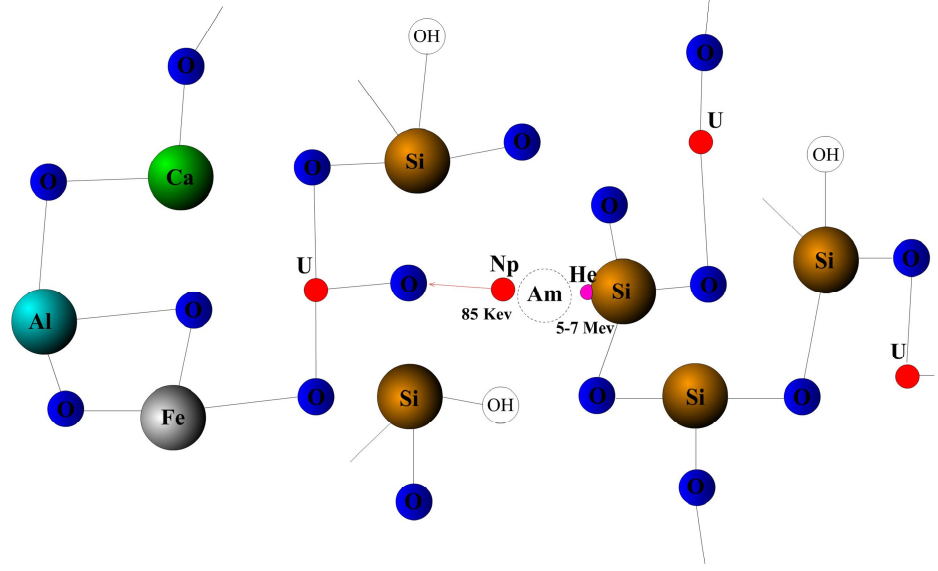}
  \caption{(Colour online) Scheme of the chemical composition of LFCM with the act of decay: $^{241}\mathrm{Am} \rightarrow \alpha + ^{237}\mathrm{Np}$.\label{Rys2}}
\end{figure}

 This model does not take into account essential radiation,
 neutron,
 electronic processes and other characteristics that are necessary for calculation and theoretical evaluations of electrical conductivity,
 thermal conductivity,
 and mechanical stability of LFCM.
 The loss of moisture in LFCM obviously has significant consequences,
 since water molecules are moderators and absorbers of neutrons.
 Non-equilibrium processes of energy transfer of particles,
 or entire clusters,
 in LFCM largely determine their stability and destruction processes.
 Undoubtedly,
 one of the important ones is the study of diffusion processes of isotopes of uranium,
 plutonium,
 americium and curium in ionic or neutral forms in LFCM.
 First of all,
 this is due to the fact that in LFCM the main sources of $\alpha$-activity and spontaneous fission neutrons are $^{238-240}\mathrm{Pu}$,
 $^{241}\mathrm{Am}$,
 $^{242}\mathrm{Am}$ and $^{242}\mathrm{Cm}$, $^{244}\mathrm{Cm}$ isotopes.

 Figure~\ref{Rys2} schematically shows the chemical composition of LFCM, which were formed by melting the content of nuclear fuel with different reactor materials.
 As a result,
 the LFCM includes chemical elements that are products of the decay of $^{235}\mathrm{U}$ in RBMK-1000 and products of neutron capture by uranium $^{238}\mathrm{U}$, which leads to plutonium isotopes $^{239}\mathrm{Pu}$, $^{240}\mathrm{Pu}$, $^{241}\mathrm{Pu}$ via neptunium isotopes $^{239}\mathrm{Np}$, as well as due to the decay of $^{241}\mathrm{Pu}$ to the isotope $^{241}\mathrm{Am}$, and then as a result of the capture of neutrons $^{241}\mathrm{Am}$ to $^{242}\mathrm{Am}$ and curium isotopes $^{242}\mathrm{Cm}$, $^{244}\mathrm{Cm}$. $^{241}\mathrm{Am}$ is the most active source of $\alpha$ - particles. As a result of its decay, an isotope of neptunium is formed ($^{241}\mathrm{Am} \rightarrow \alpha + ^{237}\mathrm{Np}$). Thermopeaks \cite{Yukhnovsky200212} are formed in the region of $^{241}\mathrm{Am}$ decay, while the interaction of the emitted $\alpha$-particles ($5-7$\,MeV) with B, O, Na, Mg, Fe, Al, Si and other atoms that are part of the LFCM is accompanied by reactions $(\alpha, n)$ and additionally generates a flow of neutrons in LFCM, which are often not taken into account in calculations. In particular,
 even earlier in works~\cite{NCPKamerton1995,NCPKamerton1995b} estimates of reaction contributions ($\alpha,n$) were made for individual samples of LFCM from sub-reactor rooms 305/2 and 304/3,
 which showed that their contribution to the rate of neutron generation in LFCM reached almost 50 percent.
 It is necessary to maintain that the rate of neutron generation due to ($\alpha, n$) reactions on light chemical elements has increased over time and will continue to increase.
 This increase is caused primarily by the accumulation of Am (as a result of the decay of Pu),
 as an intensive source of particles.
 The characteristics of neutron fluxes in LFCM rooms 305/2 and 304/3 were previously studied in works.
 In particular,
 when examining the values of separation densities,
 cadmium ratios and spectral indices for LFCM in room 304/3,
 significant differences between experimental and calculated data were observed.
 Obviously,
 this can be explained by the fact that the calculation models did not take into account the features of the composition,
 structure,
 intensity of neutron sources and the processes of radionuclide transport in LFCM.
 Such problems require the use of modern (new) approaches to the description of transport processes in porous systems.
 In this regard,
 it is important to note that in the studies of non-equilibrium processes of particle transfer
 (in particular, atoms, molecules, ions) in porous systems,
 significant success has been achieved using the generalized equations of diffusion,
 Navier--Stokes in fractional derivatives with space and time multifractality~\cite{Sahimi1998213,Korosak20071,Hobbie2007,Compte19977277,Metzler20001,Hilfer19951475,
 Hilfer20003914,Hilfer2000,Kosztolowicz2005041105,Kosztolowicz2015P10021,Bisquert20002287,Bisquert2001112,Kosztolowicz2009055004,
 Tarasov2010,Pyanylo201484,Kostrobij2016163,Kostrobij201963,Kostrobij2015154,Grygorchak2017185501,Kostrobij20184099}. Obviously,
 LFCM as porous systems are characterized by spatial fractality,
 which can significantly affect the processes of particle and energy transfer.
 Therefore,
 in our opinion, the generalized Cattaneo diffusion equations in fractional derivatives~\cite{Kostrobij2016163,Kostrobij20184099} for ions,
 taking into account fractality,
 can be used to describe transport processes during the interaction of aqueous solutions with LFCM.
 These equations have the following form:
 \begin{equation}\label{eq:2.1961}
  \tau_{a}\frac{\partial^{2}}{\partial t^{2}} n_{a}(\vec{r};t)
  +
  \frac{\partial}{\partial t} n_{a}(\vec{r};t)
  =
  {}_{0}D_{t}^{1-\xi}\sum_{b}\int \rd\mu_{\alpha}(\vec{r}')D^{\alpha}_{\vec{r}}\cdot
  \overline{D}_{q}^{ab}(\vec{r},\vec{r}')\cdot D^{\alpha}_{\vec{r}'}
  \beta\big[\gamma _{b} (\vec{r}';t)+Z_{b} e\varphi(\vec{r}';t)\big],
 \end{equation}
 where $n_{a}(\vec{r};t)$ is the non-equilibrium average value of the ion number density,
 $\gamma_{a}(\vec{r};t)$ is the non-equilibrium value of the local chemical potential of ions with valence $Z_{a}$ of type $a$;
 $e$ is the electron charge,
 $\beta =\frac{1}{k_{\mathrm{B}}T}$,
 $k_{\mathrm{B}}$ is the Boltzmann constant,
 $T$ is the equilibrium temperature,
 $\rd\mu_{\alpha}(\vec{r})=\frac{|\vec{r}|^{\alpha}}{\Gamma (\alpha)}\rd\vec{r}$, $\Gamma (\alpha)$ is the Gamma function,
 $0<\alpha<1$,
 \begin{equation}
  D^{\alpha}_{x} f(x)
  =
  \frac{1}{\Gamma (n-\alpha)}
  \int^{x}_{0} \frac{f^{(n)}(z)}{(x-z)^{\alpha +1-n}} \rd z
 \end{equation}
 is the Caputo fractal derivative~\cite{Mainardi1997,Caputo1971134,Oldham2006,Samko1993},
 ${n-1<\alpha<n}$,
 $f^{(n)}(z)=\frac{\rd^{n}}{\rd z^{n}}f(z)$ with properties:
 $D^{\alpha}_{x_{j}}1=0$ and $D^{\alpha}_{x_{j}}x_{l}=0$,
 $(j\neq l)$,
 \begin{equation}
  _{0}D^{\xi}_{t}f(t)
  =
  \frac{1}{\Gamma (\xi)}\frac{\rd}{\rd t}
  \int^{t}_{0} \frac{f(\tau)}{(t-\tau)^{1-\xi}} \rd\tau,\quad 0<\xi \leqslant 1
 \end{equation}
 is the Riemann--Liouville fractal derivative with respect to time.
 Equation~(\ref{eq:2.1961}) is a generalized Cattaneo--type equation in Gibbs statistics ($q=1$) with temporal multifractality and spatial fractality.
 The kernel $\overline{D}_{q}^{ab}(\vec{r},\vec{r}')$, according to the assumption made in paper \cite{Kostrobij20184099}, can be approximately represented as
 \[
 \overline{D}_{q}^{ab}(\vec{r},\vec{r}')=\int \rd\omega\, W^{-1}_{a}(\ri\omega)D_{q}^{ab}(\vec{r},\vec{r}';\ri\omega+\varepsilon),
 \]
 where
 \[
 W_{a}(\ri\omega)=\frac{(\ri\omega)^{1-\xi}}{1+\ri\omega \tau_{a}}, \quad
 0<\xi\leqslant 1,
 \]
 where the introduced relaxation time $\tau_{a}$ characterizes the charge carriers transport processes
in the system, $\omega$ is the frequency and $D_{q}^{ab}(\vec{r},\vec{r}';\ri\omega+\varepsilon)$ is the Laplace representation ($z=\ri\omega+\varepsilon$, $\varepsilon \rightarrow +0$) of the time correlation function of the microscopic density of ion velocities of the sorts $a$ and $b$:
\[
D_{q}^{ab}(\vec{r},\vec{r}';t,t')=\langle \vec{v}_{a}(\vec{r})T(t,t')\vec{v}_{b}(\vec{r}') \rangle^{t}_{\alpha,\rm rel}
\]
is the generalized mutual diffusion coefficient of the corresponding sorts $a$ and $b$ within the Renyi statistics, where $\vec{v}_{a}(\vec{r})=\sum_{j=1}^{N_{a}}\vec{v}_{j}\delta (\vec{r}-\vec{r}_{j})$ is the microscopic ion flux density of sort $a$,
$\langle (...) \rangle^{t}_{\alpha,\rm rel}$ is the averaging operation with the relevant distribution function corresponding to the Renyi statistics \cite{Kostrobij20184099}.
 It is important to note that the right-hand side of the equation~(\ref{eq:2.1961}) includes the fractional derivative of the scalar potential of the electromagnetic field
 $\frac{\partial^{\alpha}}{\partial\vec{r}^{'\alpha}}\beta Z_{b}e\varphi(\vec{r}';t)$,
 which indicates the need to consider the system of Maxwell's equations in fractional derivatives for a system with spatial fractality for a complete description of ion transport processes in such an environment.
 This approach in combination with electrochemical impedance was successfully applied to the description of lithium ion intercalation processes in clathrate systems~\cite{Grygorchak2017185501,Kostrobij20184099}.

 \section{Conclusions}

 Since there is a decrease in humidity in the modern conditions of NSC-SO, based on the calculations,
 it can be concluded that the diffusion of uranium and other tranuranic atoms also decreases.
 And that is obviously good.
 However,
 in the processes of dust suppression in the SO with liquid mixtures of a localizing polymer coating and a liquid with gadolinium (previously it was from 45 tons to 90 tons per year),
 there is a problem of disposal of radioactive water, which obviously interacts in certain places with LFCM,
 in particular in the lower rooms.
 Therefore,
 in connection with the continuous $\alpha-\beta-\gamma$---irradiation of the aqueous solution from the surface of the LFCM,
 tracks of densely ionized plasma are constantly formed in the near-surface layer,
 consisting of electrons,
 ions  $\mathrm{H}^{+}$, $\mathrm{OH}^{-}$ are products of the decomposition of water molecules,
 as well as ions present in the aqueous  solution $\mathrm{Na}^{+}$, $\mathrm{CO}_{3}^{2-}$, $\mathrm{HCO}_{3}^{-}$, $\mathrm{UO}_{2}^{2+}$, $\mathrm{PuO}_{2}^{2+}$, $\mathrm{Cs}^{+}$, $\mathrm{Sr}^{2+}$ and others.
 Complex diffusion,
 adsorption,
 oxidation-reduction processes take place in this near-surface region, accompanied by the processes of hydrolysis by $\mathrm{H}^{+}$, $\mathrm{OH}^{-}$, $\mathrm{Na}^{+}$, $\mathrm{Ca}^{2+}$ ions and polymerization of the silicon-oxygen network, leaching of ions from the surface of the LFCM.
 After a time of $10^{-12}\div10^{-8}$\,s,
 the electrons and part of the $\mathrm{H}^{+}$, $\mathrm{OH}^{-}$ are ions that leave the track areas,
 going into the solution or into the LFCM matrix.
 In addition,
 radiolysis products $\mathrm{HO}_{2}$, $\mathrm{H}_{2}\mathrm{O}_{2}$, $\mathrm{O}_{2}^{-}$ are formed,
 which take an active part in oxidation-reduction processes with the participation of $\mathrm{UO}_{2}^{2+}$, $\mathrm{PuO}_{2}^{2+}$, $\mathrm{CO}_{3}^{2-}$, $\mathrm{HCO}_{3}^{-}$, $\mathrm{Na}^{+}$,
 with the formation of charged and neutral complexes.
 Together with $e_{aq}^{-}$, $\mathrm{H}^{+}$, $\mathrm{OH}^{-}$ that escaped recombination,
 they begin to diffuse from the tracks,
 filling the area of the solution that is not irradiated,
 while creating significant gradients of particle densities and the electromagnetic field.
 On the basis of our diffusion model,
 it is possible to estimate the leaching process of ions from the porous matrix as a result of the internal hydrolysis reaction front penetrating it~(\ref{eq.02}).
 At the same time,
 leaching of ions increases under the condition $v_{h} \gg (D/t)^{1/2}$.
 At the same time,
 as shown by the calculations in the near-surface region of the volume phase ``solution -- surface of LFCM'',
 heterogeneous ion diffusion coefficients,
 in particular for $\mathrm{UO}_{2}^{2+}$, $\mathrm{Cs}^{+}$ have a region of fast diffusion.
 Here it is important to note that as a result of the negative charge of the LFCM surface,
 hydrogen ions formed in the radiolysis processes in the surface region are quickly electrostatically attracted to the LFCM surface and thereby activate the ion exchange mechanisms for the release of radionuclides from the LFCM matrix into the aqueous solution.

To conduct better theoretical calculations within the framework of this model or based on (\ref{eq:2.1961}) for further prediction of the behavior of LFCM in NSC-SO, it is necessary to study the reaction constants~(\ref{eq.01})--(\ref{eq.011}) for alkali-carbonate solutions in the near-surface region of the bulk phase ``solution -- LFCM surface''.
It would be important to conduct such mathematical modeling in combination with electrochemical impedance methods to study electrochemical reactions,
 diffusion processes,
 and adsorption in the near-surface region of the volume phase ``solution -- surface of LFCM''.
 In particular,
 electrochemical impedance methods were used in studies of corrosion processes of materials in interaction with highly active aqueous solutions~\cite{Bellanger199783}.

\section*{Appendix}

\[
\bar{\lambda}_{2}^{\alpha\alpha}\vec{r})=\frac{1}{3m_{\alpha}}\left\{\frac{\partial ^{2}}{\partial \vec{r}^{2}}U_{\alpha}(\vec{r})
+\sum_{\gamma}\rho_{\gamma}
\int \left[\frac{\partial ^{2}}{\partial \vec{r}^{2}}\Phi_{\alpha\gamma}(\vec{r},\vec{r}')\right]\frac{F_{2}^{\alpha\gamma}(\vec{r},\vec{r}')}{f_{1}^{\alpha}(\vec{r})}\rd\vec{r}'\right\}
\]
\[
+\frac{5}{3m_{\alpha}}\left\{\frac{\partial ^{2}}{\partial \vec{r}^{2}}U_{\alpha}(\vec{r})
+\sum_{\gamma}\rho_{\gamma}
\int \left[\left(\frac{\partial ^{2}}{\partial \vec{r}^{2}}\Phi_{\alpha\gamma}(\vec{r},\vec{r}')\right)\frac{F_{2}^{\alpha\gamma}(\vec{r},\vec{r}')} {f_{1}^{\alpha}(\vec{r})}
+ \frac{\partial }{\partial \vec{r}}\Phi_{\alpha\gamma}(\vec{r},\vec{r}')\frac{\partial }{\partial \vec{r}}\frac{F_{2}^{\alpha\gamma}(\vec{r},\vec{r}')}{f_{1}^{\alpha}(\vec{r})}\right]\rd\vec{r}'\right\},
\]
where $U_{\alpha}(\vec{r})={\varepsilon_{\alpha}(\vec{r})}/{f_{1}^{\alpha}(\vec{r})}$, $\varepsilon_{\alpha}(\vec{r})$ is the average value of the potential energy density of ions of the sort~$\alpha$ , $\Phi_{\alpha\gamma}(\vec{r},\vec{r}')$ is the pair interaction potential between ions of the sorts $\alpha$ and $\gamma$, $f_{1}^{\alpha}(\vec{r})$ is the equilibrium unary distribution function of ions of the sort $\alpha$, $F_{2}^{\alpha\gamma}(\vec{r},\vec{r}')$ is the equilibrium pair distribution function of ions of the sorts $\alpha$ and $\gamma$. Detailed expressions for $\Phi_{\alpha\gamma}(\vec{r},\vec{r}')$, $f_{1}^{\alpha}(\vec{r})$, $F_{2}^{\alpha\gamma}(\vec{r},\vec{r}')$, taking into account the screening effects in the corresponding approximations, are given in the work~\cite{Yukhnovskii2000361}.


\newpage

\ukrainianpart

\title{Опис процесів взаємодії води та водних розчинів з паливовмісними матеріалами в новому безпечному конфайнменті об’єкта ``Укриття»''}
\author{М. В. Токарчук\refaddr{label1,label2},
        Б. М. Маркович\refaddr{label2}, О. С. Захар'яш\refaddr{label2}, О. Л. Іванків\refaddr{label1}, С. М. Мохняк\refaddr{label2}}
\addresses{
\addr{label1} Інститут фізики конденсованих систем НАН України, 79011 Львів, вул. Свєнціцького, 1
\addr{label2} Національний університет ``Львівська політехніка'', 79013, м. Львів, вул. С. Бандери, 12
}
%
%
%

\makeukrtitle

\begin{abstract}
\tolerance=3000%
 Представлено основні механізми та умови взаємодії лавоподібних паливовмісних матеріалів (ЛПВМ) з атмосферою,
 водою та водними розчинами. Проаналізовано механізми руйнування поверхні ЛПВМ,
 в тому числі іонообмінні процеси, гідроліз, розчинення, окислення та ін.
 Розраховано коефіцієнти неоднорідної дифузії іонів UO$_{2}^{2+}$, Cs$^{+}$ на межі розділу ``водний розчин радіоактивних елементів -- ЛПВМ''. Проаналізовано процеси взаємодифузії та швидкість проникнення у пористе середовище фронту реакції внутрішнього гідролізу кремнієво-кисневої сітки під час взаємодії з водним розчином, коли
 процеси реакції та адсорбції приховані у швидкостях внутрішнього гідролізу.

\keywords дифузія, пористе середовище, гідроліз, реакції, адсорбція.

\end{abstract}

\end{document}